\documentclass[aps,prl,twocolumn,letterpaper,showpacs]{revtex4-1}
\usepackage[utf8]{inputenc}
\usepackage{graphicx,amsmath}
\usepackage{amssymb}
\usepackage{amsthm}
\usepackage{multirow}

\DeclareMathAlphabet{\mathitb}{OT1}{cmr}{bx}{sl}

\begin{document}
\title{Time reversal symmetric topological exciton condensate in bilayer HgTe quantum wells}
\author{Jan Carl Budich$^{1,2,3}$}
\author{Bj{\"o}rn Trauzettel$^{4}$}
\author{Paolo Michetti$^{4,5}$}

\affiliation{$^1$Department of Physics, Stockholm University, Se-106 91 Stockholm, Sweden;\\
$^2$Institute for Quantum Optics and Quantum Information,
Austrian Academy of Sciences, 6020 Innsbruck, Austria;\\
$^3$Institute for Theoretical Physics, University of Innsbruck, 6020 Innsbruck,Austria;\\
$^4$Institute for Theoretical Physics and Astrophysics,
 University of W$\ddot{u}$rzburg, 97074 W$\ddot{u}$rzburg, Germany;\\
 $^5$Institute of Theoretical Physics, Technische Universität Dresden, 01062 Dresden, Germany}
\date{\today}
\begin{abstract}
We investigate a bilayer system of critical HgTe quantum wells each featuring a spin-degenerate pair of massless Dirac fermions.
In the presence of an electrostatic inter-layer Coulomb coupling, we determine the exciton condensate order parameter of the system self-consistently.
Calculating the bulk topological $\mathbb Z_2$~invariant of the resulting mean field Hamiltonian,
we discover a novel time reversal symmetric topological exciton condensate state coined the helical topological exciton condensate.
We argue that this phase can exist for experimentally relevant parameters.
Interestingly, due to its multi-band nature, the present bilayer model exhibits a nontrivial interplay between spontaneous symmetry breaking
and topology: Depending on which symmetry the condensate order parameter spontaneously picks in combined orbital and spin space,
stable minima in the free energy corresponding to both trivial and nontrivial gapped states can be found.
\end{abstract}
\maketitle

{\emph{Introduction. }}
Incented by the theoretical prediction \cite{KaneMele2005a,KaneMele2005b,BHZ2006} and experimental observation \cite{koenig2007} of the time reversal
symmetric quantum spin Hall (QSH) state in HgTe quantum wells (QWs), the study of topological states of matter (TSM)
which can be understood on the basis of quadratic translation invariant Hamiltonians
has been among the most rapidly growing research fields in physics in recent years \cite{HasanKane,XLReview2010,TSMReview}.
The historically first TSM in this sense is the quantum anomalous Hall (QAH) state \cite{QAH}, a lattice version of
the integer quantum Hall effect \cite{Klitzing1980,Laughlin1981,TKNN1982}.
An exhaustive classification of all possible TSM in the ten Altland-Zirnbauer (AZ) symmetry classes \cite{AltlandZirnbauer}
of insulators and mean field superconductors has been achieved by different means in Refs. \cite{Schnyder2008,KitaevPeriodic,RyuLudwig}.
In some analogy to the condensation of Cooper pairs in a superconductor (SC), attractive interaction between electrons and holes in crystals can lead to macroscopically coherent quantum phenomena, relating to the formation of an exciton condensate (EC)~\cite{blatt1962,moskalenko1962,keldysh1964}. However, from a viewpoint of topological classification, there is a crucial difference between SCs and ECs: While the Nambu description of an SC leads to an algebraic constraint that can formally be viewed as a particle hole symmetry, the mean field Hamiltonian of an EC has the form of an ordinary particle number conserving insulator.

From a more practical point of view, bilayer semiconductor nanostructures are in 
principle ideally suited for the formation of an  EC~\cite{lozovik1976}. 
The reason is that, in these systems, long exciton lifetimes are expected since the spatial separation of particles and holes (which are paired to form
\emph{indirect} excitons \cite{lozovik1976, snoke2011}) is particularly suitable for minimizing detrimental interband 
processes~\cite{shevchenko1994}.
Two phenomena relating to the formation of ECs are commonly distinguished in the literature \cite{lozovik1976}.
On the one hand bilayer systems have been employed to study the condensation of indirect excitons rooted in the separation of an optically pumped electron-hole plasma.
On the other hand an (equilibrium) excitonic insulator phase can be realized \cite{jerome}
where the spontaneous instability towards electron-hole pairing gives rise to a gapped spectrum,
accompanied by an electronic longitudinal sound wave excitation mode. Here, we are concerned with the latter excitonic insulator class of ECs.
\begin{figure}
\includegraphics[width=8.5cm]{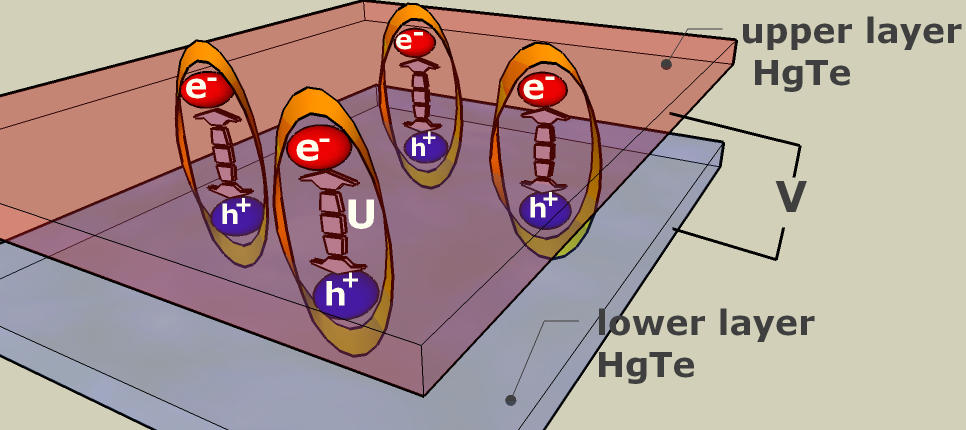}
\caption{(Color online) Schematic of a bilayer HgTe system with spatially separated two-dimensional electron and hole gases. Electrostatic coupling of the layers (indicated by double arrows) leads to electron-hole pairing between the layers (indicated by orange strips). Under certain circumstances, specified in the text, a topologically non-trivial EC phase emerges. The bilayer system can be externally tuned through an inter-layer potential bias $V$.}
\label{fig:sketch}
\end{figure}

It is fair to say that conclusive evidence of an EC in a bilayer quantum well system 
is still elusive although promising experimental observations, in particular in the quantum Hall regime, have been made in the past twenty years \cite{mur1994,spi2000,spi2001,yoo2010,fin2011,hua2012,nandi2012}.

Recent research directions in the field of ECs in bilayer Dirac systems include the study of bilayer graphene~\cite{lozovik2008,zhang2008,min2008,kharitonov2008}, interfacing surfaces of three-dimensional (3D)
topological insulators (TIs)~\cite{TECMoore,HaoTEC,ChoMooreMagnetic,tilahun2011,Efimkin2012}, and double quantum wells embedded in semiconductor heterostructures  with strong spin-orbit interaction~\cite{can2009,HaoChiralTEC}.
An overview over the experimental state of the field is provided by Refs.~\cite{dasgupta2011,snoke2011}.\\

Pioneering work towards devising a topologically nontrivial EC has been reported in Ref. \cite{TECMoore}
employing two layers of surface states of 3D TIs \cite{Fu3DTI,Moore2007,Hsieh2008}.
Each of these surface states consists of a single Dirac cone with a chiral spin structure that is chosen opposite for the two layers.
Such a chiral Dirac cone along with the exciton induced gap brings about half of the $k$-space winding needed
for a QAH state with non-vanishing Chern number \cite{TKNN1982}.
Since the QAH state relies on the breaking of time reversal symmetry (TRS), this winding cannot be completed by taking into account the higher energy modes of the TRS preserving 3D TI state \cite{ChoMooreMagnetic}.
Hence, when considering a full translation invariant model of the 3D TI, the total $k$-space topology of the EC becomes trivial.
To cure this issue, a TRS breaking extension of the model reported in Ref. \cite{TECMoore} that relies on the coupling to a ferromagnetic layer has been
presented and identified as a topological exciton condensate (TEC) in the QAH universality
class \cite{HaoTEC,ChoMooreMagnetic,HaoChiralTEC}, i.e., in the AZ class A in 2D.
Like the QAH state, the TEC is characterized by its first Chern number.
The experimental realization of such a magnetically insulating phase is very challenging. This is also reflected in the
fact that first experimental signatures of the QAH state have only been reported very recently \cite{qahscience}, although the QAH state is seemingly the conceptually simplest TSM.\\

{\emph{Main results. }}
In this Letter, we report the discovery of a novel time reversal symmetric topological exciton condensate state in a system of two coupled HgTe quantum wells (CQWs) --
the helical topological exciton condensate (HTEC) (see Fig. \ref{fig:sketch} for a schematic).
In a single HgTe QW with a finite inverted bandgap, the QSH phase has been theoretically
predicted \cite{BHZ2006} and experimentally discovered \cite{koenig2007}.
In contrast, in our setting, the single-particle spectrum of each of the quantum wells features a spin-degenerate pair of massless Dirac cones.
The coupling between the quantum wells is only due to electrostatic Coulomb interaction.
In formal analogy to the insulating QSH phase, the HTEC is characterized by a nontrivial $\mathbb Z_2$~topological
invariant \cite{KaneMele2005b,FuPump,Prodanz2}, but the presence of a gap is solely due to the formation of an EC.
Solving the self-consistent mean-field equations for the CQWs, we observe an intriguing interplay between spontaneous symmetry
breaking and the topological state of the system rooted in its multi orbital nature.
More precisely, we find stable self consistent solutions for both trivial and non-trivial values of the topological $\mathbb Z_2$~invariant
depending on the matrix structure of the order parameter in combined spin and orbital space.
Interestingly, the ordinary Dirac mass parameter $M$~ in the individual layers competes with the EC gap.
The former can be experimentally tuned by changing the thickness of the layers away from the critical thickness of $d_c=6.4$~nm \cite{BHZ2006,koenig2007}. This provides a knob to change the topological phase of the system.
We argue that the parameter regime we investigate is in the right ballpark for state of the art experiments on HgTe quantum wells. In a ribbon geometry, the HTEC is shown to feature gapless helical edge states.\\

{\emph{Model for the HTEC. }}
The low energy sector of a single massless ($M=0$) HgTe QW is described by the Bernevig, Hughes, and Zhang (BHZ) model~\cite{BHZ2006}.
\begin{equation}
H_{BHZ}(k) = A k_x \sigma_x s_z + \left(A k_y \sigma_y - B k^2 \sigma_z - D k^2 \sigma_z  \right) s_0,
\label{eqn:bhz}
\end{equation}
with $\sigma$ and $s$~ denoting the Pauli matrices associated with the band pseudospin and the real spin degrees of freedom, respectively.
Importantly, massless HgTe QWs have already been experimentally realized~\cite{Buettner2011}.
Starting from this model Hamiltonian (\ref{eqn:bhz}) for the individual quantum wells, we now consider a CQW system.
The separation $t$~between the two QWs is assumed to be sufficiently large to exclude single-particle tunneling which is detrimental for the formation of an EC. The dependence of tunneling matrix elements on $t$~has been microscopically analyzed for this system in Ref.~\cite{PaoloBilayer}.
Thus, the EC in the CQW is described, up to an overall constant, by the following mean-field Hamiltonian \cite{Note1}:
\begin{eqnarray}
\hat H_{MF} &=& \sum_\mathitb k \hat\Psi_\mathitb k^\dag ~\mathbb{H}(\mathitb k)~ \hat\Psi_\mathitb k \; ,\\
\mathbb H(k) &=& \tau_0 H_{BHZ} (k) - \frac{V}{2} \tau_z + \mathbb H_I(k)
\label{eq:H}
\end{eqnarray}
with the spinor $\hat \Psi$ = $\Big(\hat a_{\mathitb k}^{(E_1,\uparrow)}$, $\hat a_{\mathitb k}^{(E_1,\downarrow)}$,
$\hat a_{\mathitb k}^{(H_1,\uparrow)}$, $\hat a_{\mathitb k}^{(H_1,\downarrow)}$,
$\hat b_{\mathitb k}^{(E_1,\uparrow)}$, $\hat b_{\mathitb k}^{(E_1,\downarrow)}$,
$\hat b_{\mathitb k}^{(H_1,\uparrow)}$, $\hat b_{\mathitb k}^{(H_1,\downarrow)}\Big)$,
where $\hat a$ and $\hat b$ are the annihilation operators for the electron and hole layer, respectively.
We also introduce $\tau$, the Pauli matrices in the electron-hole (or layer) pseudo spin space, and the interlayer potential bias $V$,
tunable by gating the CQW structure.
The coupling between the layers is then described by
\begin{eqnarray}
[\mathbb H_I(k)]_{\alpha,\beta} &=& \sum_\mathitb{k'}
V_{|\mathitb k-\mathitb{k'}|} \langle \hat \Psi_\mathitb k^\dag(\alpha) \hat \Psi_\mathitb k(\beta)  \rangle \; ,
\label{eq:coupling}
\\
\langle \hat \Psi_\mathitb k^\dag(\beta) \hat \Psi_\mathitb k(\alpha)  \rangle &=& \sum_i^8 \left[\mathbb{U}^\dag\right]_{i,\beta} \left[\mathbb{U}\right]_{\alpha,i}
f(E_\mathitb k^{(i)}),
\label{eq:coupling2}
\end{eqnarray}
with $\alpha\in[1,4]$ and $\beta\in[5,8]$, or the other way around (only pairing interaction terms), with the Coulomb
inter-layer interaction $V_q= \frac{2\pi e^2}{\epsilon q}e^{-qt}$ and the effective dielectric constant $\epsilon$.
$\mathbb U$ is the unitary transformation that diagonalizes $\mathbb H$, while $f(E)$ accounts for the
Fermi-Dirac occupation of the quasi-particle eigenstate $E_\mathitb k^{(i)}$.
We focus on $s$-wave pairing, i.e., we assume the order parameter, and hence $\mathbb H_I$,
to have an isotropic $k$-dependence. Additionally, we neglect the intralayer Coulomb interactions because it is not essential for the formation of the EC.

The main conceptual advantage of our present model compared to the 3D TI setup reported in Ref. \cite{TECMoore}
is that the chiral winding of the surface Dirac cones is now in band pseudo-spin space instead of real-spin space.
Therefore, the combination of band-bending at higher energies and an exciton condensate induced gap can complete this winding
to a topologically nontrivial band structure even in the presence of TRS.

Starting from an initial seed for $\mathbb H_I(k)$, we calculate the self-consistent solution for the EC
order parameter $[\mathbb H_I(k)]_{\alpha,\beta}$~appearing in Eq.(\ref{eq:coupling}) which can be decomposed as
\begin{equation}
 \mathbb H_I(\mathitb k)  = \sum_{i=1}^{16} U_i(k) \mathbb D_i,
\end{equation}
where $U_i$ is the mean-field pairing potential with matrix structure $\mathbb D_i$.
$\mathbb D_i$ is one of the $16$ linearly independent matrices that are off-diagonal in the layer
pseudospin (containing $\tau_x$ or $\tau_y$), and commute with the time reversal
operator $\mathcal T = i \tau_0 \sigma_0 s_y K$~with $K$~denoting the complex conjugation.

{\emph{Topological classification and results.  }}
Within this space of $16$ real parameters, we determine the order parameter self consistently
and characterize the resulting mean-field band structure topologically.
The present two dimensional model system obeys TRS with $\mathcal T^2=-1$~and is thus in the AZ class AII.
It is hence characterized by the $\mathbb Z_2$ invariant $\mathcal V= 0,1$ \cite{KaneMele2005b,FuPump,Prodanz2},
which we calculate by employing the manifestly gauge invariant method due to Prodan \cite{Prodanz2}.
This method uses the gauge invariant adiabatic connection introduced by Kato \cite{Kato1950} and
provides a numerically straightforward recipe for the calculation of $\mathcal V$~in the absence of any additional symmetry.

In Table~\ref{tab:orderparameters}, we summarize the possible EC phases resulting from our mean field calculation for an
inter-well potential bias $V=0$ and a background screening constant $\epsilon=5$.
Depending on the initial seed, we self-consistently observe the establishment  of
either one or two coexisting species of order parameters of the form  $\mathbb D=\tau_i \sigma_j s_k$ (denoted by $ijk$ in Table~\ref{tab:orderparameters}),
that determine the nature of the EC.
In Table~\ref{tab:orderparameters}, we list all the possible matrix structures of the order parameter $ijk$~ which lead to self-consistent
gapped EC phases together with the strength of the self-consistent pairing potential $U$,
the size of the energy gap, as well as the topological invariant $\mathcal V=0,1$ where  $\mathcal V= 1$ characterizes the nontrivial HTEC phase.
Pairs of compatible allowed order parameter structures (for example xz0 and yzz),
which differ only in their spin part, are grouped together, because they realize ECs of identical topological properties with the same EC gap.
This behavior is related to the spin degeneracy of the underlying BHZ model.
We choose this value of $\epsilon$ to demonstrate that the EC gap would be quite large (several meV)
even in the case $V=0$, if a spacing layer with such a low dielectric constant could be experimentally realized.
Therefore, it is desirable to find a rather small $\epsilon$ material as a spacing layer.
However, for current experiments on HgTe QWs, CdTe with $\epsilon=10$~is used as a spacing layer.

\begin{table}[h]
\begin{tabular}{|c|c|c|c|}
\hline
$ijk$~ &   $U$ (meV)~ & gap (meV)~ & ~$\mathcal V$\\
\hline
xz0 & \multirow{2}{*}{2.27}     & \multirow{2}{*}{4.55}   &  \multirow{2}{*}{1}   \\%
 yzz & & &\\%
\hline
 yxx & \multirow{2}{*}{2.40}       & \multirow{2}{*}{4.80}  & \multirow{2}{*}{0}     \\
 yxy &       &   &      \\%
 \hline
\end{tabular}
\caption{
Left column: Matrix structure of order parameter $\mathbb D =\tau_i\sigma_js_k$ realizing an EC phase,
denoted by $ijk$. Central left column: Self-consistent value of the pairing field.
Central right column: Self-consistent value of the EC gap.
Right column: Value of the topological $\mathbb Z_2$~ invariant $\mathcal V$.\\
$A=250$~meV~nm, $B=-1000$~meV~nm$^{\rm2}$, $D=M=0$, $\epsilon=5$.
}
\label{tab:orderparameters}
\end{table}

\begin{figure}
\includegraphics[width=0.8\linewidth]{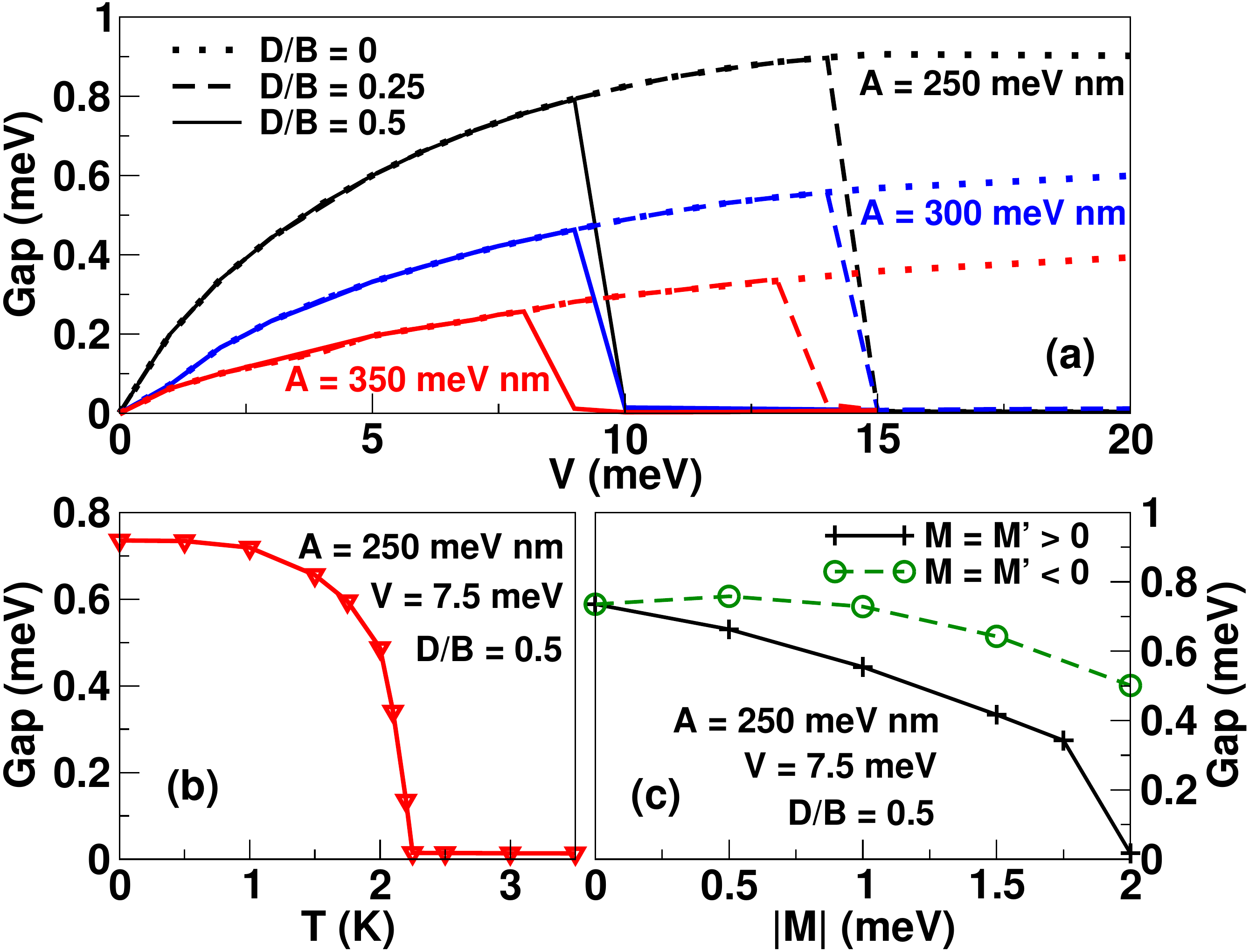}
\caption{(Color online) (a) Gap of the HTEC phase $xz0$ as a function of the inter-well potential bias $V$ for $\epsilon=10$ at zero temperature
for different values of $A$ and $D/B$. $M=0$ and $B=-1000$~meV~nm$^{\rm2}$.
In (b) and (c) we closely analyze the case $A=250$~meV~nm, $V=7.5$~meV and $D/B=0.5$, with $B=-1000$~meV~nm$^{\rm2}$.
(b) Gap of the HTEC phase $xz0$ as a function of the temperature for $M=0$.
(c) Gap of the HTEC phase $xz0$ as a function of the Dirac masses $M=M'$ of electron and hole layers.
}
\label{fig:Vdep}
\end{figure}

In Fig.~\ref{fig:Vdep}(a), we analyze the full $V$ dependence of the EC gap of
the non-trivial HTEC phase with the pairing structure $xz0$, for the more realistic choice $\epsilon=10$.
Curves are presented, corresponding to different values of the Dirac term $A$ and of the particle-hole symmetry breaking term $D$ in the BHZ model,
while keeping fixed the parameter $B=-1000$~meV~nm$^{\rm 2}$.
In a purely Dirac system ($B=D=0$), we can define an effective fine-structure constant $\alpha=e^2/(\epsilon A)$ that determines the importance of Coulomb interactions~\cite{graphene_review}. Hence, the smaller the parameter $A$, the more important are Coulomb interactions.
This trend still applies to some extent to the BHZ model whose screening properties have recently been analyzed within random phase approximation~\cite{jurgens}. 
Therefore, the increase of the EC gap with increasing $\alpha$ can be understood from this perspective, {\it cf.} Fig.~\ref{fig:Vdep}(a).
This figure illustrates that, for a particle-hole symmetric spectrum $D=0$, the EC gap monotonously increases with $V$ because the density of electron and hole states that interact with each other also monotonously increases.
At the next level of complication, a finite $D$ parameter leads to a mismatch of electron and hole Fermi surfaces due to the band bending terms in the BHZ model,
which becomes significant for wavevectors on the order of $A/(B+D)$, eventually leading to the suppression of inter-well coherence.
Indeed, for a fixed $A$ parameter, finite $D$ curves perfectly follow the $D=0$ one for sufficiently small $V$,
but coherence suddenly disappears at a cut-off $V_c$, which is a monotonic decreasing function of $D/B$.
We note that realistic values for the $A$ parameters may depend upon the QW thickness as well as on the chemical composition of the spacing material between
the two HgTe QWs.
A small $A$ parameter is beneficial to the observation of an EC phase in this system.
In the literature, the estimates for the $A$ parameter vary in a range between $250$ and $400$~meV~nm, for which the maximal expected EC gap value
varies from about $1$~meV to $0.1$~meV, respectively.
Hence, the predicted phenomenon could be within reach of state of the art experiments on HgTe quantum wells.
In Fig.~\ref{fig:Vdep}(b), we show the temperature dependence of the mean-field EC energy gap for $A=250$~meV and $D/B=0.5$ at $V=7.5$~meV.
The HTEC phase is seen to be stable up to $2$~K. Therefore, we conclude that it may be feasible to see this phase in low-temperature experiments.
In Fig.~\ref{fig:Vdep}(c), we analyze the effect of possible (for simplicity, identical) finite Dirac masses $M=M'$ of the particle and hole layers, which lead to
energy gaps $2|M| \ll V$ in the dispersion relation of each of the two QWs.
The EC gap is clearly reduced for positive $M$, while it is less sensitive to negative $M$ (it even slightly increases for small negative Dirac masses).

\begin{figure}
\includegraphics[width=0.49\linewidth]{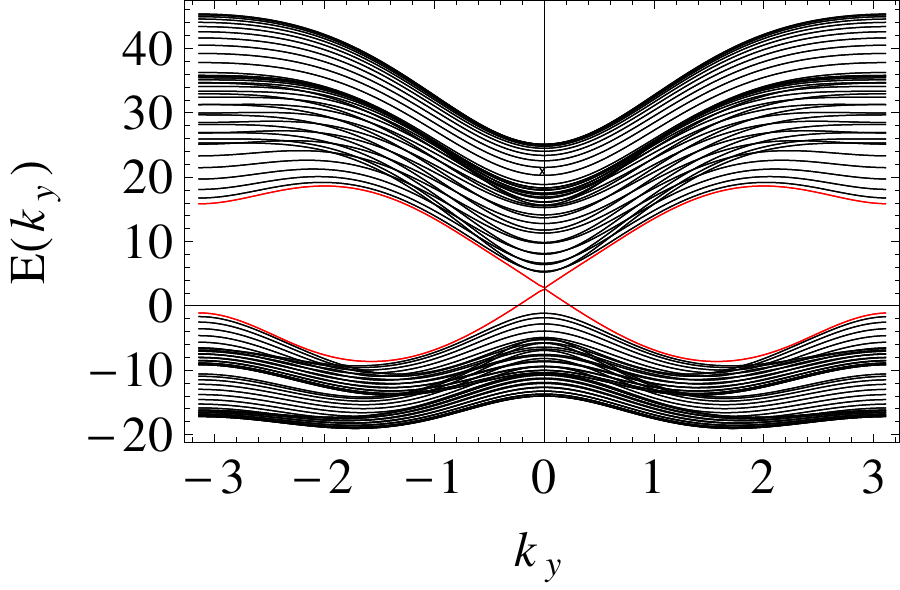}
\includegraphics[width=0.49\linewidth]{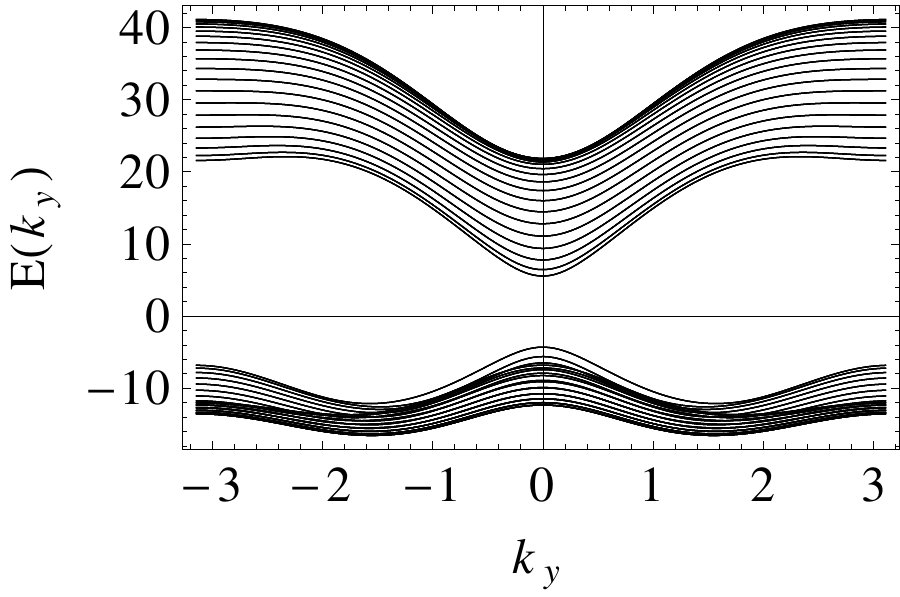}
\caption{(Color online) Ribbon spectra for a width of $20$~ sites in $x$-direction and a lattice regularization with lattice constant $a=20\,nm$. Energies are measured in $meV$, momenta in units of $a^{-1}$. Left: Topologically nontrivial order parameter $\tau_x \sigma_z s_0$.
Spin degenerate edge modes at opposite edges are marked in red. Right: Topologically trivial order parameter $\tau_y \sigma_x  s_y$ where no edge states appear. $A=275$~meV~nm, $B=-1300$~meV~nm$^2$, $D=-730$~meV~nm$^2$, $M=0$~in both panels.}
\label{fig:edgemodes}
\end{figure}

In Fig. \ref{fig:edgemodes}, we show the spectrum of a ribbon in the HTEC phase ($xz0$) and in a trivial EC phase ($yxy$), respectively.
The topologically non-trivial EC exhibits fermionic helical edge states which cross the EC gap. While, similar to a SC,
the single particle energy gap is filled by a condensate (here an EC which gives rise to transport effects relating to superfluidity),
the edge states do not have the nature of Majorana modes as in a SC but are similar to those of an ordinary QSH state. This phenomenology
corroborates the picture of a topological exciton condensate as a state of matter in between a conventional topological insulator and a superconductor.

{\emph{Concluding remarks. }} We investigated a critical bilayer system of HgTe quantum wells in which an exciton condensate gap was shown to arise
from electrostatic inter layer coupling.
Most interestingly, we found stable self consistent order parameters corresponding to both a topologically trivial
and non-trivial time reversal symmetric exciton condensate state.
For experimentally relevant microscopic parameters, the exciton condensate gap was shown to reach about 1 meV
which makes our predictions experimentally observable at low temperatures.

{\emph{Acknowledgments. }}
We would like to thank Stefan Juergens for interesting discussions. JCB acknowledges financial support from the Swedish Research Council (VR). BT and PM are financially supported by the DFG (FOR 1483 and SPP 1666) as well as the Helmholtz Foundation (VITI).

{\emph{Note added. }} While preparing the submission of this manuscript, we became aware of a related work on the interplay of exciton condensation and quantum spin Hall effect in InAs/GaSb bilayer structures \cite{pik2013}.


\begin{thebibliography}{42}
\expandafter\ifx\csname natexlab\endcsname\relax\def\natexlab#1{#1}\fi
\expandafter\ifx\csname bibnamefont\endcsname\relax
  \def\bibnamefont#1{#1}\fi
\expandafter\ifx\csname bibfnamefont\endcsname\relax
  \def\bibfnamefont#1{#1}\fi
\expandafter\ifx\csname citenamefont\endcsname\relax
  \def\citenamefont#1{#1}\fi
\expandafter\ifx\csname url\endcsname\relax
  \def\url#1{\texttt{#1}}\fi
\expandafter\ifx\csname urlprefix\endcsname\relax\def\urlprefix{URL }\fi
\providecommand{\bibinfo}[2]{#2}
\providecommand{\eprint}[2][]{\url{#2}}

\bibitem[{\citenamefont{Kane and Mele}(2005{\natexlab{a}})}]{KaneMele2005a}
\bibinfo{author}{\bibfnamefont{C.~L.} \bibnamefont{Kane}} \bibnamefont{and}
  \bibinfo{author}{\bibfnamefont{E.~J.} \bibnamefont{Mele}},
  \bibinfo{journal}{Phys. Rev. Lett.} \textbf{\bibinfo{volume}{95}},
  \bibinfo{pages}{226801} (\bibinfo{year}{2005}{\natexlab{a}}).

\bibitem[{\citenamefont{Kane and Mele}(2005{\natexlab{b}})}]{KaneMele2005b}
\bibinfo{author}{\bibfnamefont{C.~L.} \bibnamefont{Kane}} \bibnamefont{and}
  \bibinfo{author}{\bibfnamefont{E.~J.} \bibnamefont{Mele}},
  \bibinfo{journal}{Phys. Rev. Lett.} \textbf{\bibinfo{volume}{95}},
  \bibinfo{pages}{146802} (\bibinfo{year}{2005}{\natexlab{b}}).

\bibitem[{\citenamefont{Bernevig et~al.}(2006)\citenamefont{Bernevig, Hughes,
  and Zhang}}]{BHZ2006}
\bibinfo{author}{\bibfnamefont{B.~A.} \bibnamefont{Bernevig}},
  \bibinfo{author}{\bibfnamefont{T.~L.} \bibnamefont{Hughes}},
  \bibnamefont{and} \bibinfo{author}{\bibfnamefont{S.-C.} \bibnamefont{Zhang}},
  \bibinfo{journal}{Science} \textbf{\bibinfo{volume}{314}},
  \bibinfo{pages}{1757} (\bibinfo{year}{2006}).

\bibitem[{\citenamefont{K{\"o}nig et~al.}(2007)\citenamefont{K{\"o}nig,
  Wiedmann, Br\"une, Roth, Buhmann, Molenkamp, Qi, and Zhang}}]{koenig2007}
\bibinfo{author}{\bibfnamefont{M.}~\bibnamefont{K{\"o}nig}},
  \bibinfo{author}{\bibfnamefont{S.}~\bibnamefont{Wiedmann}},
  \bibinfo{author}{\bibfnamefont{C.}~\bibnamefont{Br\"une}},
  \bibinfo{author}{\bibfnamefont{A.}~\bibnamefont{Roth}},
  \bibinfo{author}{\bibfnamefont{H.}~\bibnamefont{Buhmann}},
  \bibinfo{author}{\bibfnamefont{L. W.}~\bibnamefont{Molenkamp}},
  \bibinfo{author}{\bibfnamefont{X.-L.} \bibnamefont{Qi}}, \bibnamefont{and}
  \bibinfo{author}{\bibfnamefont{S.-C.} \bibnamefont{Zhang}},
  \bibinfo{journal}{Science} \textbf{\bibinfo{volume}{318}},
  \bibinfo{pages}{766} (\bibinfo{year}{2007}).

\bibitem[{\citenamefont{Hasan and Kane}(2010)}]{HasanKane}
\bibinfo{author}{\bibfnamefont{M.~Z.} \bibnamefont{Hasan}} \bibnamefont{and}
  \bibinfo{author}{\bibfnamefont{C.~L.} \bibnamefont{Kane}},
  \bibinfo{journal}{Rev. Mod. Phys.} \textbf{\bibinfo{volume}{82}},
  \bibinfo{pages}{3045} (\bibinfo{year}{2010}).

\bibitem[{\citenamefont{Qi and Zhang}(2011)}]{XLReview2010}
\bibinfo{author}{\bibfnamefont{X.-L.} \bibnamefont{Qi}} \bibnamefont{and}
  \bibinfo{author}{\bibfnamefont{S.-C.} \bibnamefont{Zhang}},
  \bibinfo{journal}{Rev. Mod. Phys.} \textbf{\bibinfo{volume}{83}},
  \bibinfo{pages}{1057} (\bibinfo{year}{2011}).

\bibitem[{\citenamefont{{Budich} and {Trauzettel}}(2013)}]{TSMReview}
\bibinfo{author}{\bibfnamefont{J.~C.} \bibnamefont{{Budich}}} \bibnamefont{and}
  \bibinfo{author}{\bibfnamefont{B.}~\bibnamefont{{Trauzettel}}},
  \bibinfo{journal}{Physica Status Solidi Rapid Research Letters}
  \textbf{\bibinfo{volume}{7}}, \bibinfo{pages}{109} (\bibinfo{year}{2013}).

\bibitem[{\citenamefont{Haldane}(1988)}]{QAH}
\bibinfo{author}{\bibfnamefont{F.~D.~M.} \bibnamefont{Haldane}},
  \bibinfo{journal}{Phys. Rev. Lett.} \textbf{\bibinfo{volume}{61}},
  \bibinfo{pages}{2015} (\bibinfo{year}{1988}).

\bibitem[{\citenamefont{Klitzing et~al.}(1980)\citenamefont{Klitzing, Dorda,
  and Pepper}}]{Klitzing1980}
\bibinfo{author}{\bibfnamefont{K.~v.} \bibnamefont{Klitzing}},
  \bibinfo{author}{\bibfnamefont{G.}~\bibnamefont{Dorda}}, \bibnamefont{and}
  \bibinfo{author}{\bibfnamefont{M.}~\bibnamefont{Pepper}},
  \bibinfo{journal}{Phys. Rev. Lett.} \textbf{\bibinfo{volume}{45}},
  \bibinfo{pages}{494} (\bibinfo{year}{1980}).

\bibitem[{\citenamefont{Laughlin}(1981)}]{Laughlin1981}
\bibinfo{author}{\bibfnamefont{R.~B.} \bibnamefont{Laughlin}},
  \bibinfo{journal}{Phys. Rev. B} \textbf{\bibinfo{volume}{23}},
  \bibinfo{pages}{5632(R)} (\bibinfo{year}{1981}).

\bibitem[{\citenamefont{Thouless et~al.}(1982)\citenamefont{Thouless, Kohmoto,
  Nightingale, and den Nijs}}]{TKNN1982}
\bibinfo{author}{\bibfnamefont{D.~J.} \bibnamefont{Thouless}},
  \bibinfo{author}{\bibfnamefont{M.}~\bibnamefont{Kohmoto}},
  \bibinfo{author}{\bibfnamefont{M.~P.} \bibnamefont{Nightingale}},
  \bibnamefont{and} \bibinfo{author}{\bibfnamefont{M.}~\bibnamefont{den Nijs}},
  \bibinfo{journal}{Phys. Rev. Lett.} \textbf{\bibinfo{volume}{49}},
  \bibinfo{pages}{405} (\bibinfo{year}{1982}).

\bibitem[{\citenamefont{Altland and Zirnbauer}(1997)}]{AltlandZirnbauer}
\bibinfo{author}{\bibfnamefont{A.}~\bibnamefont{Altland}} \bibnamefont{and}
  \bibinfo{author}{\bibfnamefont{M.~R.} \bibnamefont{Zirnbauer}},
  \bibinfo{journal}{Phys. Rev. B} \textbf{\bibinfo{volume}{55}},
  \bibinfo{pages}{1142} (\bibinfo{year}{1997}).

\bibitem[{\citenamefont{Schnyder et~al.}(2008)\citenamefont{Schnyder, Ryu,
  Furusaki, and Ludwig}}]{Schnyder2008}
\bibinfo{author}{\bibfnamefont{A.~P.} \bibnamefont{Schnyder}},
  \bibinfo{author}{\bibfnamefont{S.}~\bibnamefont{Ryu}},
  \bibinfo{author}{\bibfnamefont{A.}~\bibnamefont{Furusaki}}, \bibnamefont{and}
  \bibinfo{author}{\bibfnamefont{A.~W.~W.} \bibnamefont{Ludwig}},
  \bibinfo{journal}{Phys. Rev. B} \textbf{\bibinfo{volume}{78}},
  \bibinfo{pages}{195125} (\bibinfo{year}{2008}).

\bibitem[{\citenamefont{Kitaev}(2009)}]{KitaevPeriodic}
\bibinfo{author}{\bibfnamefont{A.}~\bibnamefont{Kitaev}}, \bibinfo{journal}{AIP
  Conference Proceedings} \textbf{\bibinfo{volume}{1134}}, \bibinfo{pages}{22}
  (\bibinfo{year}{2009}).

\bibitem[{\citenamefont{{Ryu} et~al.}(2010)\citenamefont{{Ryu}, {Schnyder},
  {Furusaki}, and {Ludwig}}}]{RyuLudwig}
\bibinfo{author}{\bibfnamefont{S.}~\bibnamefont{{Ryu}}},
  \bibinfo{author}{\bibfnamefont{A.~P.} \bibnamefont{{Schnyder}}},
  \bibinfo{author}{\bibfnamefont{A.}~\bibnamefont{{Furusaki}}},
  \bibnamefont{and} \bibinfo{author}{\bibfnamefont{A.~W.~W.}
  \bibnamefont{{Ludwig}}}, \bibinfo{journal}{New Journal of Physics}
  \textbf{\bibinfo{volume}{12}}, \bibinfo{pages}{065010}
  (\bibinfo{year}{2010}).

\bibitem[{\citenamefont{Blatt et~al.}(1962)\citenamefont{Blatt, B{\"o}er, and
  Brandt}}]{blatt1962}
\bibinfo{author}{\bibfnamefont{J.~M.} \bibnamefont{Blatt}},
  \bibinfo{author}{\bibfnamefont{K.~W.} \bibnamefont{B{\"o}er}},
  \bibnamefont{and} \bibinfo{author}{\bibfnamefont{W.}~\bibnamefont{Brandt}},
  \bibinfo{journal}{Phys. Rev.} \textbf{\bibinfo{volume}{126}},
  \bibinfo{pages}{1691} (\bibinfo{year}{1962}).

\bibitem[{\citenamefont{Moskalenko}(1962)}]{moskalenko1962}
\bibinfo{author}{\bibfnamefont{S.}~\bibnamefont{Moskalenko}},
  \bibinfo{journal}{Fiz. Tverd. Tela} \textbf{\bibinfo{volume}{4}},
  \bibinfo{pages}{276} (\bibinfo{year}{1962}).

\bibitem[{\citenamefont{Keldysh and Kopaev}(1964)}]{keldysh1964}
\bibinfo{author}{\bibfnamefont{L.~V.} \bibnamefont{Keldysh}} \bibnamefont{and}
  \bibinfo{author}{\bibfnamefont{Y.~V.}~\bibnamefont{Kopaev}},
  \bibinfo{journal}{Fiz. Tverd. Tela} \textbf{\bibinfo{volume}{6}},
  \bibinfo{pages}{2791} (\bibinfo{year}{1964}).

\bibitem[{\citenamefont{Lozovik and Yudson}(1976)}]{lozovik1976}
\bibinfo{author}{\bibfnamefont{Y.~E.} \bibnamefont{Lozovik}} \bibnamefont{and}
  \bibinfo{author}{\bibfnamefont{V.~I.} \bibnamefont{Yudson}},
  \bibinfo{journal}{Zh. Eksp. Teor. Fiz.} \textbf{\bibinfo{volume}{71}},
  \bibinfo{pages}{738} (\bibinfo{year}{1976}).


  \bibitem[{\citenamefont{Snoke}(2011)}]{snoke2011}
\bibinfo{author}{\bibfnamefont{D.~W.} \bibnamefont{Snoke}},
  \bibinfo{journal}{Adv. in Condens. Matter Phys.} \textbf{\bibinfo{volume}{2011}}, \bibinfo{pages}{938609}
  (\bibinfo{year}{2011}).
  

\bibitem[{\citenamefont{Shevchenko}(1994)}]{shevchenko1994}
\bibinfo{author}{\bibfnamefont{S.~I.} \bibnamefont{Shevchenko}},
  \bibinfo{journal}{Phys. Rev. Lett.} \textbf{\bibinfo{volume}{72}},
  \bibinfo{pages}{3242} (\bibinfo{year}{1994}).

  

\bibitem{jerome}
D.~Jérome, T.~M.~Rice and W.~Kohn, Phys.~Rev. {\bf 158}, 158 (1967).
  
\bibitem{mur1994}
S.~Q.~Murphy~{\it et al.}, Phys. Rev. Lett. {\bf 72}, 728 (1994).

\bibitem{spi2000}
I.~B.~Spielman~{\it et al.}, Phys. Rev. Lett. {\bf 84}, 5808 (2000).

\bibitem{spi2001}
I.~B.~Spielman~{\it et al.}, Phys. Rev. Lett. {\bf 87}, 036803 (2001).

\bibitem{yoo2010}
Y.~Yoon~{\it et al.}, Phys. Rev. Lett. {\bf 104}, 116802 (2010).

\bibitem{fin2011}
A.~D.~K.~Finck~{\it et al.}, Phys. Rev. Lett. {\bf 106}, 236807 (2011).

\bibitem{hua2012}
X.~Huang~{\it et al.}, Phys. Rev. Lett. {\bf 109}, 156802 (2012).

\bibitem[{\citenamefont{{Nandi~{\it et al.}}}(2012)}]{nandi2012}
\bibinfo{author}{\bibfnamefont{D.}~\bibnamefont{{Nandi~{\it et al.}}}},
  \bibinfo{journal}{Nature} \textbf{\bibinfo{volume}{488}},
  \bibinfo{pages}{481} (\bibinfo{year}{2012}).

\bibitem[{\citenamefont{Lozovik and Sokolik}(2008)}]{lozovik2008}
\bibinfo{author}{\bibfnamefont{Y.~E.} \bibnamefont{Lozovik}} \bibnamefont{and}
  \bibinfo{author}{\bibfnamefont{A.~A.} \bibnamefont{Sokolik}},
  \bibinfo{journal}{JETP Lett.} \textbf{\bibinfo{volume}{87}},
  \bibinfo{pages}{55} (\bibinfo{year}{2008}).

\bibitem[{\citenamefont{Zhang and Joglekar}(2008)}]{zhang2008}
\bibinfo{author}{\bibfnamefont{C.-H.} \bibnamefont{Zhang}} \bibnamefont{and}
  \bibinfo{author}{\bibfnamefont{Y.~N.} \bibnamefont{Joglekar}},
  \bibinfo{journal}{Phys. Rev. B} \textbf{\bibinfo{volume}{77}},
  \bibinfo{pages}{233405} (\bibinfo{year}{2008}).


\bibitem[{\citenamefont{Min et~al.}(2008)\citenamefont{Min, Bistritzer, Su, and
  MacDonald}}]{min2008}
\bibinfo{author}{\bibfnamefont{H.}~\bibnamefont{Min}},
  \bibinfo{author}{\bibfnamefont{R.}~\bibnamefont{Bistritzer}},
  \bibinfo{author}{\bibfnamefont{J.-J.} \bibnamefont{Su}}, \bibnamefont{and}
  \bibinfo{author}{\bibfnamefont{A.~H.} \bibnamefont{MacDonald}},
  \bibinfo{journal}{Phys. Rev. B} \textbf{\bibinfo{volume}{78}},
  \bibinfo{pages}{121401(R)} (\bibinfo{year}{2008}).

\bibitem[{\citenamefont{Kharitonov and Efetov}(2008)}]{kharitonov2008}
\bibinfo{author}{\bibfnamefont{M.~Y.} \bibnamefont{Kharitonov}}
  \bibnamefont{and} \bibinfo{author}{\bibfnamefont{B.~B.}
  \bibnamefont{Efetov}}, \bibinfo{journal}{Phys. Rev. B}
  \textbf{\bibinfo{volume}{78}}, \bibinfo{pages}{241401(R)}
  (\bibinfo{year}{2008}).

\bibitem[{\citenamefont{Seradjeh et~al.}(2009)\citenamefont{Seradjeh, Moore,
  and Franz}}]{TECMoore}
\bibinfo{author}{\bibfnamefont{B.}~\bibnamefont{Seradjeh}},
  \bibinfo{author}{\bibfnamefont{J.~E.} \bibnamefont{Moore}}, \bibnamefont{and}
  \bibinfo{author}{\bibfnamefont{M.}~\bibnamefont{Franz}},
  \bibinfo{journal}{Phys. Rev. Lett.} \textbf{\bibinfo{volume}{103}},
  \bibinfo{pages}{066402} (\bibinfo{year}{2009}).

\bibitem[{\citenamefont{Hao et~al.}(2011)\citenamefont{Hao, Zhang, and
  Wang}}]{HaoTEC}
\bibinfo{author}{\bibfnamefont{N.}~\bibnamefont{Hao}},
  \bibinfo{author}{\bibfnamefont{P.}~\bibnamefont{Zhang}}, \bibnamefont{and}
  \bibinfo{author}{\bibfnamefont{Y.}~\bibnamefont{Wang}},
  \bibinfo{journal}{Phys. Rev. B} \textbf{\bibinfo{volume}{84}},
  \bibinfo{pages}{155447} (\bibinfo{year}{2011}).

\bibitem[{\citenamefont{Cho and Moore}(2011)}]{ChoMooreMagnetic}
\bibinfo{author}{\bibfnamefont{G.~Y.} \bibnamefont{Cho}} \bibnamefont{and}
  \bibinfo{author}{\bibfnamefont{J.~E.} \bibnamefont{Moore}},
  \bibinfo{journal}{Phys. Rev. B} \textbf{\bibinfo{volume}{84}},
  \bibinfo{pages}{165101} (\bibinfo{year}{2011}).

\bibitem[{\citenamefont{Tilahun et~al.}(2011)\citenamefont{Tilahun, Lee,
  Hankiewicz, and MacDonald}}]{tilahun2011}
\bibinfo{author}{\bibfnamefont{D.}~\bibnamefont{Tilahun}},
  \bibinfo{author}{\bibfnamefont{B.}~\bibnamefont{Lee}},
  \bibinfo{author}{\bibfnamefont{E.~M.} \bibnamefont{Hankiewicz}},
  \bibnamefont{and}
  \bibinfo{author}{\bibfnamefont{A.}~\bibnamefont{MacDonald}},
  \bibinfo{journal}{Phys. Rev. Lett.} \textbf{\bibinfo{volume}{107}},
  \bibinfo{pages}{246401} (\bibinfo{year}{2011}).



 \bibitem{Efimkin2012}
 D.K. Efimkin, Yu. E. Lozovik, and A.A. Sokolik, Phys. Rev. B {\bf 86}, 115436 (2012).


\bibitem[{\citenamefont{Can and Hakio\ifmmode~\breve{g}\else
  \u{g}\fi{}lu}(2009)}]{can2009}
\bibinfo{author}{\bibfnamefont{M.~A.} \bibnamefont{Can}} \bibnamefont{and}
  \bibinfo{author}{\bibfnamefont{T.}~\bibnamefont{Hakio\ifmmode~\breve{g}\else
  \u{g}\fi{}lu}}, \bibinfo{journal}{Phys. Rev. Lett.}
  \textbf{\bibinfo{volume}{103}}, \bibinfo{pages}{086404}
  (\bibinfo{year}{2009}).

\bibitem[{\citenamefont{Hao et~al.}(2010)\citenamefont{Hao, Zhang, Li, Wang,
  Zhang, and Wang}}]{HaoChiralTEC}
\bibinfo{author}{\bibfnamefont{N.}~\bibnamefont{Hao}},
  \bibinfo{author}{\bibfnamefont{P.}~\bibnamefont{Zhang}},
  \bibinfo{author}{\bibfnamefont{J.}~\bibnamefont{Li}},
  \bibinfo{author}{\bibfnamefont{Z.}~\bibnamefont{Wang}},
  \bibinfo{author}{\bibfnamefont{W.}~\bibnamefont{Zhang}}, \bibnamefont{and}
  \bibinfo{author}{\bibfnamefont{Y.}~\bibnamefont{Wang}},
  \bibinfo{journal}{Phys. Rev. B} \textbf{\bibinfo{volume}{82}},
  \bibinfo{pages}{195324} (\bibinfo{year}{2010}).

\bibitem{dasgupta2011}
K. Das Gupta, A. F. Croxall, J. Waldie, C. A. Nicoll, H. E. Beere, I. Farrer, D. A. Ritchie and M. Pepper, 
\bibinfo{journal}{Adv. in Condens. Matter Phys.} \textbf{\bibinfo{volume}{2011}}, \bibinfo{pages}{727958}
(\bibinfo{year}{2011}).

\bibitem[{\citenamefont{Fu et~al.}(2007)\citenamefont{Fu, Kane, and
  Mele}}]{Fu3DTI}
\bibinfo{author}{\bibfnamefont{L.}~\bibnamefont{Fu}},
  \bibinfo{author}{\bibfnamefont{C.~L.} \bibnamefont{Kane}}, \bibnamefont{and}
  \bibinfo{author}{\bibfnamefont{E.~J.} \bibnamefont{Mele}},
  \bibinfo{journal}{Phys. Rev. Lett.} \textbf{\bibinfo{volume}{98}},
  \bibinfo{pages}{106803} (\bibinfo{year}{2007}).

\bibitem[{\citenamefont{Moore and Balents}(2007)}]{Moore2007}
\bibinfo{author}{\bibfnamefont{J.~E.} \bibnamefont{Moore}} \bibnamefont{and}
  \bibinfo{author}{\bibfnamefont{L.}~\bibnamefont{Balents}},
  \bibinfo{journal}{Phys. Rev. B} \textbf{\bibinfo{volume}{75}},
  \bibinfo{pages}{121306(R)} (\bibinfo{year}{2007}).

\bibitem[{\citenamefont{Hsieh et~al.}(2009)\citenamefont{Hsieh, Qian, Wray,
  Xia, Hor, Cava, and Hasan}}]{Hsieh2008}
\bibinfo{author}{\bibfnamefont{D.}~\bibnamefont{Hsieh}},
  \bibinfo{author}{\bibfnamefont{D.}~\bibnamefont{Qian}},
  \bibinfo{author}{\bibfnamefont{L.}~\bibnamefont{Wray}},
  \bibinfo{author}{\bibfnamefont{Y.}~\bibnamefont{Xia}},
  \bibinfo{author}{\bibfnamefont{Y.~S.} \bibnamefont{Hor}},
  \bibinfo{author}{\bibfnamefont{R.~J.} \bibnamefont{Cava}}, \bibnamefont{and}
  \bibinfo{author}{\bibfnamefont{M.~Z.} \bibnamefont{Hasan}},
  \bibinfo{journal}{Nature} \textbf{\bibinfo{volume}{452}},
  \bibinfo{pages}{970} (\bibinfo{year}{2009}).

  \bibitem[{\citenamefont{Chang et~al.}(2013)\citenamefont{Chang, Zhang, Feng,
  Shen, Zhang, Guo, Li, Ou, Wei, Wang et~al.}}]{qahscience}
\bibinfo{author}{\bibfnamefont{C.-Z.} \bibnamefont{Chang}},
  \bibinfo{author}{\bibfnamefont{J.}~\bibnamefont{Zhang}},
  \bibinfo{author}{\bibfnamefont{X.}~\bibnamefont{Feng}},
  \bibinfo{author}{\bibfnamefont{J.}~\bibnamefont{Shen}},
  \bibinfo{author}{\bibfnamefont{Z.}~\bibnamefont{Zhang}},
  \bibinfo{author}{\bibfnamefont{M.}~\bibnamefont{Guo}},
  \bibinfo{author}{\bibfnamefont{K.}~\bibnamefont{Li}},
  \bibinfo{author}{\bibfnamefont{Y.}~\bibnamefont{Ou}},
  \bibinfo{author}{\bibfnamefont{P.}~\bibnamefont{Wei}},
  \bibinfo{author}{\bibfnamefont{L.-L.} \bibnamefont{Wang}},
  \bibnamefont{et~al.}, \bibinfo{journal}{Science}
  \textbf{\bibinfo{volume}{340}}, \bibinfo{pages}{167} (\bibinfo{year}{2013}).

\bibitem{Buettner2011}
B. B{\"u}ttner, C. X. Liu, G. Tkachov, E. G. Novik, C. Br{\"u}ne, H. Buhmann, E. M. Hankiewicz, P. Recher, B. Trauzettel, S. C. Zhang, and L. W. Molenkamp, Nature Phys. {\bf 7}, 418 (2011).

\bibitem[{\citenamefont{Michetti et~al.}(2012)\citenamefont{Michetti, Budich,
  Novik, and Recher}}]{PaoloBilayer}
\bibinfo{author}{\bibfnamefont{P.}~\bibnamefont{Michetti}},
  \bibinfo{author}{\bibfnamefont{J.~C.} \bibnamefont{Budich}},
  \bibinfo{author}{\bibfnamefont{E.~G.} \bibnamefont{Novik}}, \bibnamefont{and}
  \bibinfo{author}{\bibfnamefont{P.}~\bibnamefont{Recher}},
  \bibinfo{journal}{Phys. Rev. B} \textbf{\bibinfo{volume}{85}},
  \bibinfo{pages}{125309} (\bibinfo{year}{2012}).

\bibitem[{Note1()}]{Note1}
 {Similar to Ref. \cite {BHZ2006}, we neglect here bulk
  inversion symmetry breaking terms which would couple the two spin blocks
  within the individual QWs. Our results do not crucially rely on this
  simplification which however makes the physical picture of the HTEC much more
  transparent and intuitive. The HTEC phase is robust against the adiabatic
  switching-on of such terms}.

\bibitem[{\citenamefont{Fu and Kane}(2006)}]{FuPump}
\bibinfo{author}{\bibfnamefont{L.}~\bibnamefont{Fu}} \bibnamefont{and}
  \bibinfo{author}{\bibfnamefont{C.~L.} \bibnamefont{Kane}},
  \bibinfo{journal}{Phys. Rev. B} \textbf{\bibinfo{volume}{74}},
  \bibinfo{pages}{195312} (\bibinfo{year}{2006}).

\bibitem[{\citenamefont{Prodan}(2011)}]{Prodanz2}
\bibinfo{author}{\bibfnamefont{E.}~\bibnamefont{Prodan}},
  \bibinfo{journal}{Phys. Rev. B} \textbf{\bibinfo{volume}{83}},
  \bibinfo{pages}{235115} (\bibinfo{year}{2011}).

\bibitem[{\citenamefont{Kato}(1950)}]{Kato1950}
\bibinfo{author}{\bibfnamefont{T.}~\bibnamefont{Kato}},
  \bibinfo{journal}{Journal of the Physical Society of Japan}
  \textbf{\bibinfo{volume}{5}}, \bibinfo{pages}{435} (\bibinfo{year}{1950}).

\bibitem{graphene_review}
 V.~N.~Kotov, B.~Uchoa, V.~M.~Pereira, F.~Guinea and A.~H.~Castro~Neto, Rev. Mod. Phys. {\bf 84}, 1067 (2012).

\bibitem{jurgens}
S.~J\"urgens, P.~Michetti and B.~Trauzettel,
 \bibinfo{journal}{Phys. Rev. Lett.} \textbf{\bibinfo{volume}{112}},
  \bibinfo{pages}{076804} (\bibinfo{year}{2014}).


  \bibitem{pik2013}
D.~I.~Pikulin and T~Hyart, arXiv:1311.1111 (2013).


\end{thebibliography}
\end{document}